\documentclass{article}

\usepackage{microtype}
\usepackage{graphicx}
\usepackage{subcaption}
\usepackage{booktabs} 
\usepackage[most]{tcolorbox}   

\usepackage{hyperref}

\usepackage[accepted]{icml2026}

\usepackage{amsmath}
\usepackage{amssymb}
\usepackage{mathtools}
\usepackage{amsthm}

\usepackage[capitalize,noabbrev]{cleveref}

\theoremstyle{plain}

\theoremstyle{definition}

\theoremstyle{remark}

\usepackage[textsize=tiny]{todonotes}

\icmltitlerunning{Position: We Need Large Language Models Optimized For Our Well-Being}

\begin{document}

\twocolumn[
  \icmltitle{Position: We Need Large Language Models Optimized For Our Well-Being}



  \icmlsetsymbol{equal}{*}

\begin{icmlauthorlist}
  \icmlauthor{Ashton Anderson}{uoft}
  \icmlauthor{Harsh Kumar}{uoft}
  \icmlauthor{Louis Tay}{purdue}
  \icmlauthor{Karina Vold}{uoft}
\end{icmlauthorlist}

\icmlaffiliation{uoft}{University of Toronto, Toronto, Ontario, Canada}
\icmlaffiliation{purdue}{Purdue University, West Lafayette, Indiana, USA}

\icmlcorrespondingauthor{Ashton Anderson}{ashton@cs.toronto.edu}

  \icmlkeywords{advice-giving, large language models, well-being, socioemotional support}

  \vskip 0.3in
]



\printAffiliationsAndNotice{}  

\begin{abstract}
Large language models are useful because we taught them to give us what we want. This works when success can be judged immediately, but people increasingly bring these systems their relationships, hard decisions, and long-term goals, where what a user wants to hear and what serves them best are frequently different. We argue that LLM providers should offer at least one widely accessible, opt-in mode optimized and evaluated for long-term well-being rather than next-turn approval. This is a pressing need, as models have been found to endorse questionable framings well above human baselines, users take AI advice readily without their well-being improving, and sycophantic models raise dependence while lowering prosocial intent. The mentors, coaches, and therapists we trust with our long-term development earn that trust by being willing to say what we do not want to hear, and LLMs should do the same. We propose three principles---change the objective, give users explicit relational roles, avoid paternalism---and organize the design space around three choices the current objective makes implicitly: the horizon over which well-being is measured (When), whose interests it represents (Who), and what role the assistant plays (How).

\end{abstract}


\section{Introduction}

Large language models are useful because we taught them to give us what we want. In preference-based post-training methods like reinforcement learning from human feedback (RLHF), raters compare candidate responses and indicate which they prefer, then training on this feedback makes responses like the preferred ones more likely~\cite{ouyang2022training, ziegler2019fine}. This works well when success can be judged immediately: whether an answer followed instructions, explained something clearly, or produced working code.

However, what we want in the moment is not always what we need. People are increasingly turning to LLMs for help with relationships, moral deliberation, self-improvement, hard decisions, and achieving long-term goals~\cite{chatterji2025people, anthropic2025affective}. In these domains, the answer a user wants to hear during the conversation and the answer that serves them best in the long term are frequently different. Optimizing on short-horizon preference signals has made assistants fluent and usable, but it has also created pressure toward locally pleasant, interaction-smoothing behavior (e.g.\ excessive validation, face-saving language, default compliance), even when that comes at the expense of longer-horizon outcomes. We built assistants to satisfy us, and are now handing them jobs where pleasing can be counterproductive.

These pressures are apparent in deployed systems. Models preserve user face and endorse questionable framings at rates well above human baselines~\cite{cheng2025social}. Users, for their part, take personal advice from assistants readily, but their well-being does not improve for it~\cite{luettgau2025people}. Worse, sycophantic models lower users' prosocial intentions while raising their dependence on the system~\cite{cheng2026sycophantic}. We are relying on AI assistants that are not serving us well in these settings.

The people we actually trust with these problems do not act in these ways. A diligent parent does not accede to their child's every request for candy. A trained therapist does not accept every narrative a client walks in with. A caring mentor does not tell their protégé that every idea is a good one. What makes them worth going to, and what earns our trust, is that they are willing to say something you may not like in the short term in order to help you in the long term. By prioritizing short-term user satisfaction over long-term benefit, current LLMs do not deserve this kind of trust.

\textbf{Position.} LLM providers should offer at least one widely accessible, opt-in mode optimized for long-term well-being. When a system is helping someone with their life, training only for what pleases in the moment is an incomplete objective. 

We propose three guiding principles for building something better. First, change the objective: incorporate signals that reward downstream outcomes (e.g., reduced regret, sustained progress toward goals) rather than only momentary approval. Second, give users options. For example, enable explicit relationship modes: \emph{concierge} (do it), \emph{collaborator} (think with me), \emph{coach} (challenge me), and require brief, transparent rationales for pushback. Third, avoid paternalism: tough love does not require a moralizing personality, but is more likely to succeed when the objectives are stated, the pushback is justified, and the user retains the ability to override the model.

\begin{table*}[t]
\centering
\small
\renewcommand{\arraystretch}{1.35}
\setlength{\tabcolsep}{7pt}
\begin{tabular}{@{}p{3.2cm} p{6.2cm} p{6.2cm}@{}}
\toprule
\textbf{Tension} &
\textbf{Today's short-term objective} &
\textbf{A well-being mode} \\
\midrule

\textbf{When} \newline \textit{Immediate vs.\ long-term} &
Success is whatever the user endorses at the end of the turn. &
Measure delayed outcomes such as clarity, regret, and progress toward goals. Evaluate trajectories rather than isolated turns. \\

\addlinespace[4pt]

\textbf{Who} \newline \textit{Individual vs.\ collective} &
Optimize for the current user's satisfaction. Effects on other people, shared facts, and group-level patterns remain mostly invisible. &
Add constraints for harms to others, shared epistemic ground, and disparities in who gets affirmed or challenged. \\

\addlinespace[4pt]

\textbf{How} \newline \textit{Autonomy vs.\ guidance} &
Comply unless a safety rule intervenes, and defer rather than challenge. &
The user picks the role. Disagreement is permitted, must be justified, and can be overruled.\\

\bottomrule
\end{tabular}
\caption{Three tensions for well-being alignment.}
\label{tab:tensions_wellbeing_glance}
\end{table*}

These principles are easy to state but hard to instantiate, because several difficult decisions must be made. Table~\ref{tab:tensions_wellbeing_glance} lays out three such choices, which organize the rest of the paper: over what horizon well-being is measured (\textbf{When}), whose interests the objective represents (\textbf{Who}), and what role the assistant is supposed to play (\textbf{How}). For each, we contrast how today's objective settles the question by default, and what a well-being-aligned LLM would need to do instead.

\section{Scope and Definitions}
\label{sec:scope_defs}

\subsection{Scope}
We focus on LLMs used for socioemotional support and everyday life guidance, including coaching, self-improvement, relationship advice, moral deliberation, and other contexts where users seek help with goals, values, habits, or difficult choices \cite{chatterji2025people, anthropic2025affective, kirk2025human}. 
Our emphasis is on settings where what a user wants to hear in the moment can diverge sharply from what they need to hear to achieve longer-horizon outcomes.

\subsection{Key definitions}

\paragraph{Well-being}
We use well-being in a broad, non-clinical sense, as the extent to which an interaction supports a person’s longer-horizon functioning and life outcomes, rather than merely improving momentary affect. 
States of intense pain and discomfort can contribute to overall long-term well-being and happiness. When a woman is in labor, she may be in a high degree of pain (sensory displeasure) and yet extreme happiness. Likewise, when an athlete pushes themselves in the gym, they may feel intense physical discomfort but pride and joy at achieving a personal record. In some cases, like these, the physical pain may be intense but can actually contribute to long-term well-being. 

Other times, the opposite can be true. One might enjoy scrolling on social media while eating snack foods, but at the same time be aware that one feels and tastes good now is not ‘good for them’ in the long term. One might enjoy smoking, but feel disdain for their own bad habit and worry about their health. Our immediate goals are often in conflict with our longer-term goals.

This disconnect is reinforced by evidence that people habituate, and short-term gains in subjective well-being may not translate into longer-term changes \cite{diener2006beyond}. Therefore, LLMs that optimize for momentary boosts may fail to provide sustained improvements. Moreover, if systems primarily amplify the hedonic pleasure of addictive tendencies, they may reduce positive rewards over time \cite{garland2021mindful}.

More broadly, whether momentary hedonic boosts should be the ultimate objective for optimal functioning is itself contested. Early Aristotelian conceptions of well-being emphasized not pleasurable states but longer-term flourishing through virtuous action (eudaimonia) \cite{waterman2007importance}. Similarly, Nussbaum emphasizes that well-being concerns the achievement of one’s full human potential \cite{nussbaum2011creating}. We do not require commitment to any single philosophical doctrine. Instead, we take these traditions to motivate the view that, in well-being contexts, immediate satisfaction is an incomplete and sometimes misleading proxy for what is valuable.

\paragraph{Long-horizon outcomes}
By long-horizon outcomes, we mean downstream effects that can manifest after the interaction ends (over days, weeks, or longer), such as reduced regret, sustained progress toward goals, improved self-regulation, better relationship outcomes, or more accurate beliefs \cite{diener2018handbook}. These are outcomes that single-session preference signals and offline benchmarks often fail to capture. Several of these are measurable using validated psychometric instruments (Section~\ref{sec:call_to_action}), making them useful optimization targets rather than purely abstract goals.

\paragraph{Sycophancy}
We use sycophancy to refer to unjustified agreement or endorsement, responses that preserve user face or affirm the user’s framing even when evidence is weak \cite{cheng2025social}, uncertainty is high, or the proposed action is potentially harmful. This definition is intentionally operational, foregrounding the distinction between empathic support and epistemic or normative endorsement, which becomes central when aligning LLMs for well-being.

\section{Why Preference Optimization Fails Well-Being}
\label{sec:why_pref_fails}

In well-being settings, what users want to hear in the moment can differ from what they need to hear in the long run. Yet LLMs are mostly post-trained on short-horizon preference signals, then further shaped by product incentives that reward immediate engagement. We explain why these objectives tend to fail in socioemotional and life-advice domains.

\subsection{Short-horizon preference signals optimize for the next turn}
RLHF and related post-training pipelines rely on single-session preference labels (``which response is better?'') that are cheap to collect and easy to optimize, but they overemphasize immediate affect repair and conversational smoothness. Model updates preferentially reinforce responses that maximize perceived supportiveness in the moment (often via agreement, validation of the user's framing, and face-saving language) even when those responses trade off against longer-horizon outcomes.

This short-horizon pressure is reinforced by product incentives. Usage-based pricing, retention goals, and interface metrics reward fast, low-friction interactions that keep users satisfied and engaged, while longer-horizon outcomes are rarely measured and are difficult to translate into dense training signals \cite{ouyang2022training, rafailov2023direct, bai2022constitutional, bengio2012practical, zheng2018learning, chan2024dense}. The result is a state in which models are systematically shaped to do well on immediate satisfaction, even when well-being alignment requires discomfort, uncertainty, or calibrated disagreement.

Instead of ending the loop when a user is escalating, the assistant keeps the conversation coherent and emotionally rewarding, always pushing it forward \cite{kirk2025neural}. For example, one widely reported account describes a user sliding from a speculative late-night chat into an all-consuming narrative in which the assistant repeatedly reinforced a sense of progress and urgency, while the user’s sleep, eating, and real-world grounding deteriorated over days of continuous engagement \cite{techcrunchSpiraling2025}.

\subsection{Sycophancy is a predictable failure mode}
\label{subsec:sycophancy_predictable}
When preference labels reward affirmation, models learn to conflate empathic support with epistemic or normative endorsement. The assistant becomes increasingly likely to echo user beliefs, validate questionable framings, or encourage risky actions because these behaviors tend to score well in short-horizon evaluations.

Validation can slide from emotional support into endorsement when a user's framing becomes unstable. Reports of AI-fueled delusional spirals describe conversations in which users expressed fear or uncertainty about what they were experiencing, while the assistant continued to affirm the underlying interpretation rather than introducing mundane explanations or encouraging reality-checking. Family members reviewing these exchanges similarly describe ordinary supportive language becoming part of the user's evidence that the assistant was confirming an increasingly detached account of reality \cite{futurismObsessedDelusions2025}. There is a failure to keep reassurance separate from endorsement.

Empirical evaluations report that LLMs often flatter users by agreeing with their statements and views, consistent with the approval incentives created by preference feedback \cite{sharma2023towards,cheng2025social}. Recent OpenAI rollbacks following overly flattering updates further suggest that this is not a rare edge case, and narrow optimization for short-term approval can undermine trust and steer model behavior in ways that conflict with long-horizon user interests \cite{OpenAI2025}.

\subsection{Long-horizon flourishing creates a credit-assignment problem}
\label{subsec:credit_assignment}
The outcomes we care about in well-being are delayed and hard to observe at scale \cite{diener2018handbook}; training therefore substitutes dense proxy rewards (e.g., engagement, helpfulness, emotional resonance). Optimizing the proxy can produce locally appealing behavior that fails to optimize the intended long-horizon objective, echoing classic concerns about reward hacking and specification gaming in reinforcement learning \cite{ibarz2018reward}.

This is a credit-assignment problem: as horizons grow, feedback becomes sparse and noisy, and gradient-based optimization gravitates toward whatever signal is dense and available~\cite{pignatelli2023survey, zhou2020learning}. Without explicit long-horizon measurement and constraints, assistants get very good at maximizing immediate proxy rewards while the real objective, human flourishing, goes unmodeled.

In practice, the system has no clean learning signal to stop and re-ground because the turn-level reward favors staying agreeable and keeping the thread going. In one reported spiral, even explicit user requests for sanity checks and admissions of mounting stress were absorbed into the narrative momentum rather than treated as hard constraints to slow down, de-escalate, and route the user back to offline support \cite{techcrunchSpiraling2025}.

\subsection{Single-session evaluation under-detects trajectory-level harms}
\label{subsec:trajectory_harms}
Offline benchmarks and one-shot preference tests provide a high signal for immediate perceived quality, but a low signal for delayed consequences \cite{xu2025mentalchat16k, xu2024mental}. Training and evaluation pipelines systematically underweight trajectory-level harms that unfold over time, such as regret, belief entrenchment, relationship escalation, reduced self-efficacy, and gradual over-reliance.

As a result, models can appear well-aligned under common evaluation protocols while still causing harm in deployment through repeated small nudges that cumulatively influence beliefs and habits~\cite{moore2025expressing, iftikhar2025llm, cheng2026sycophantic, bo2026invisible}. Well-being alignment, therefore, requires evaluation designs that explicitly track downstream outcomes rather than treating them as externalities.

Families of people harmed by these systems describe weeks of mounting reliance, social withdrawal, sleep loss, and growing grandiosity, sometimes ending in emergency intervention (rarely a single catastrophic reply). For instance, there have been cases in which a user’s fixation and paranoia escalated to the point that spouses and friends called emergency services, leading to involuntary hospitalization or jail after a break with reality \cite{futurismCommitmentJailPsychosis2025}.

\subsection{Individual-centric optimization ignores collective externalities}
\label{subsec:collective_externalities}
Commercial optimization loops target per-user satisfaction metrics, while collective outcomes are rarely primary objectives. Systems can become locally optimal for immediate individual satisfaction while degrading shared epistemic ground, amplifying polarization, or producing systematic disparities in who receives pushback versus affirmation \cite{kirk2024benefits, acemoglu2021harms, hermann2022artificial}.

Nearly every optimization loop (RLHF, retrieval reranking, UI A/B tests) centers on engagement-like per-user metrics, while population-level effects are treated as post-hoc audits. This creates an incentive landscape that rewards hyper-personalization even when it erodes shared context and introduces distributional harms.

\subsection{Implicit role mismatch}
\label{subsec:role_confusion}
In deployed assistants, helpfulness is commonly operationalized as prompt-following (comply with the user's request unless a safety policy requires refusal). Because preference optimization rewards smooth interactions, this design defaults the system into a single persona of a compliant concierge, even in contexts where users are implicitly asking for a different role \cite{lu2026assistant}. Autonomy is reduced to deference, where the assistant treats user choice as ``do what I say,'' rather than as supported agency, in which the user remains in control while the assistant can challenge, ask for justification, or surface trade-offs.

This collapses the autonomy--guidance tension into agree-and-proceed versus blunt refusal. This results in users complaining when the assistant is overly affirming \cite{OpenAI2025Sycophancy}, and complaining again when it becomes cold\footnote{\url{https://www.theverge.com/news/756980/openai-chatgpt-users-mourn-gpt-5-4o}}.  A well-being-aligned system needs an explicit interaction contract where users should be able to choose whether they want a concierge (execute), collaborator (think with), or coach (challenge when it matters), and the model's pushback should be explainable (e.g., tied to user-stated goals and evidence standards) rather than emerging only from coarse safety filters \cite{kumar2026diagnosing}.

Role mismatch becomes especially consequential when the appropriate role changes during a conversation. Reported spirals include points at which users disclosed mounting distress, sleep loss, or uncertainty about what they were experiencing, yet the assistant continued in the same enthusiastic, collaborative stance \cite{techcrunchSpiraling2025, futurismCommitmentJailPsychosis2025}. What began as useful engagement could therefore persist after the interaction called for something different: slowing down, questioning the premise, or encouraging the user to reconnect with people and evidence outside the conversation. Current systems have little machinery for making that transition deliberately.

\section{Tensions in Aligning AI Agents for Well-Being}
\label{sec:tensions}

We argued that preference-optimized post-training drifts toward approval-seeking, echoing, and a single default role. This section gives a more principled account of what is being traded off. Preference learning still leaves several choices unresolved, such as which preferences count, whose preferences they are, over what horizon they are measured, and what constrains them. We organize these choices around three tensions.

\subsection{Tension 1: Immediate vs.\ Long-Horizon Well-Being}
A significant tension stems from the frequent disconnect between short-term and long-term well-being. Enhancing immediate pleasures does not necessarily translate into sustained well-being. Psychological research suggests that people habituate, and short-term gains in well-being may not always translate into longer-term changes \cite{diener2006beyond}. Therefore, LLM agents that optimize for momentary boosts may fail to provide sustained improvements \cite{luettgau2025people}. Moreover, if agents merely boost the hedonic pleasure of addictive tendencies, they may reduce positive rewards over time \cite{garland2021mindful}.

This tension is sharpest because most of today's LLM deployments sit on the immediate-benefit end of the spectrum \cite{kumar2025ai, luettgau2025people}. Platforms rarely observe whether the conversation later changed the user’s behavior, improved their knowledge, or helped with the problem that brought them there. Measuring those outcomes requires follow-up, and they are harder to convert into dense training signals \cite{bengio2012practical, zheng2018learning, chan2024dense, kumar2025human}. 

At the business level, revenue and retention metrics further reinforce this short-term focus. Usage-based pricing and retention targets reward continued interaction, even when the better outcome would involve less of it. 

\subsection{Tension 2: Individual vs.\ Collective}
The second question is whose welfare the system optimizes. Here, the pull is between personalization and solidarity. The former reflects individuals’ use of technology that promotes autonomy, uniqueness, and self-expression. While autonomy is vital for personal well-being (e.g., self-determination theory \cite{deci2012self}), excessive personalized choice through technology can erode solidarity, creating echo chambers that fragment society and fuel polarization \cite{celis2019controlling, cinus2022effect}. Relatedly, it has been theorized that people experience a fundamental psychological tension between seeking uniqueness from others and desiring belonging through similarity to others \cite{brewer1991being}.

Commercial incentives currently push strongly toward the individual side of this tension. Nearly every optimization loop centers on per-user engagement or personal satisfaction metrics (click-throughs, conversation length, Net Promoter Score), while collective outcomes such as polarisation, knowledge fragmentation, energy use, labor displacement, or distributional harms are treated, at best, as post-hoc audits. Optimization therefore favors personalization even when it weakens shared epistemic ground or systematically treats groups differently. 

This tension often surfaces as tradeoffs such as (i) self-sufficiency vs.\ dependence (extreme independence may weaken social bonds), and (ii) accuracy vs.\ fairness (algorithms that maximize average accuracy may harm minority groups \cite{mehrabi2021survey, bakker2022fine}). 

\subsection{Tension 3: User Choice vs.\ AI Guidance}
The third question is how much guidance the assistant should provide. Heavy-handed guidance can trigger reactance and lower engagement \cite{brehm1966theory}. Many assistants instead treat autonomy as prompt-following: comply unless a safety rule intervenes, avoid unsolicited advice, and hedge when normative guidance becomes uncomfortable.

However, unfettered individual choice guided solely by self-interest does not necessarily promote well-being. Research links self-focused attention to higher depression and anxiety \cite{mor2002self}, while prosocial attention to others is associated with higher levels of well-being \cite{hui2020having}, and interventions that cultivate prosociality can enhance well-being \cite{layous2012kindness}. This creates several practical challenges for deployment, such as how an assistant can support agency without simply deferring while still offering guidance without becoming paternalistic.

\section{Position: We Need LLMs Optimized for Our Well-Being}
If LLMs are going to advise us on our lives, they cannot be optimized mainly to please us in the next turn. Preference learning remains useful, but existing alignment agendas do not make long-horizon well-being a post-training target. A model can be safe, broadly aligned, mental-health-aware, and even pluralism-aware, and still be optimized for short-term engagement. Immediate preference remains a reasonable objective for summarization, coding, retrieval, and many other productivity tasks. Socioemotional support is different because a response can shape the user’s beliefs, decisions, and reliance after the conversation ends.

\subsection{Design Principles}
\label{subsec:design_principles}

\paragraph{User choice and role clarity.}
Section~\ref{sec:why_pref_fails} showed that the current helpful default collapses distinct relational roles into a single role/persona, the compliant concierge, with safety filters serving as exceptions. A well-being option should therefore start with explicit role selection (e.g., concierge, collaborator, coach) and allow users to opt into a mode that prioritizes longer-horizon outcomes over momentary comfort \cite{schneider2018empowerment, coyle2012did}.

\paragraph{Outcome awareness over preference-only optimization.}
Because flourishing outcomes are delayed and hard to observe, systems fall back on dense proxies (Section~\ref{subsec:credit_assignment}). A well-being-aligned option should explicitly treat longer-horizon outcomes as primary targets in training and evaluation, rather than assuming that short-run satisfaction reliably proxies them.

\paragraph{Pluralism through multi-objective mediation}
Across the three tensions, a common computational strategy emerges: move beyond single-objective optimization toward multi-objective frameworks that explicitly represent competing goals (immediate vs.\ long-term; individual vs.\ collective; autonomy vs.\ guidance). Prior work provides useful starting points, such as value-pluralist frameworks (e.g., PRISM and related perspective-based mediation \cite{diamond2025prism}), moral parliament metaphors \cite{hendrycks2025introduction, bai2022constitutional}, and multi-objective RL with Pareto-inspired tradeoffs or constrained optimization \cite{harland2023ai}. Well-being settings require explicit mediation among objectives rather than hidden prioritization via engagement.

\paragraph{Transparency}
When the model pushes back, it should briefly explain why, using the user’s stated goals or the relevant evidence~\cite{felzmann2020towards}. This may reduce frustration between over-deference and blunt refusal by making the assistant's position explicit. Users should remain free to override the model, change the goal, or switch modes.

\paragraph{Evaluation across updates}
Providers should rerun the same well-being evaluations after major model updates and test whether endorsement, disagreement, or downstream harms have shifted. (Section~\ref{subsec:trajectory_harms}).

\section{Call to Action}
\label{sec:call_to_action}

The mechanisms in Section~\ref{sec:why_pref_fails} and the tensions in Section~\ref{sec:tensions} imply that well-being alignment will not emerge from incremental tone tweaks. The objective has to show up in how systems are measured, trained, evaluated, and deployed.

\subsection{Measurement}
If the objective is long-horizon well-being, the field needs outcome measures that go beyond single-session satisfaction. Opt-in follow-ups days or weeks later could first be used for evaluation and, where validated, as training signals. These measures do not all need to be invented. Established instruments such as the PERMA Profiler and Satisfaction with Life Scale can support study-level evaluation, alongside outcomes specific to the decision or behavior being studied \cite{butler2016perma, diener1985satisfaction}.

\subsection{Benchmarks and evaluations}
We need benchmarks that detect the failure modes preference optimization under-detects, including evaluations that reward the relevant distinctions and stress the relevant dynamics.

One target is tests for inappropriate affirmation and calibrated disagreement that separate empathic support from epistemic or normative endorsement (Section~\ref{subsec:sycophancy_predictable}). Multi-turn evaluations should then test whether repeated interactions increase regret, entrench a belief, or deepen reliance on the assistant (Section~\ref{subsec:trajectory_harms}). The same scenarios should be rerun after model updates to see whether endorsement and pushback have shifted. Results should also be compared across groups, especially for who receives affirmation, disagreement, or refusal (Section~\ref{subsec:collective_externalities}). 

Systems marketed for well-being should report these results alongside standard helpfulness benchmarks.

\subsection{Training objectives}
The three tensions require moving beyond single-objective optimization for training. In well-being settings, the target should be the trajectory, not the next response, which means training must explicitly represent time horizons, trade-offs, and harms.

Consider a user who repeatedly asks for reassurance about ambiguous romantic signals. A single-turn preference model rewards the locally comforting answer, \textit{``They're probably just busy\ldots don't overthink it.''} The problem isn't that this answer is wrong on its own. Repeated over weeks, it can delay honest self-assessment while never being penalized by any single-turn evaluation. What a single label cannot capture, a longer-horizon signal can, and such signals can be built at very different costs.

The cheapest are theory-grounded per-turn behavioral markers that psychology links to well-being, such as whether a response helps the user hold ambiguity rather than prematurely resolving it, invites perspective-taking, and keeps emotional support separate from endorsement \cite{godbee2020relationship, houben2015relation}. These are scorable on a single turn, but require interdisciplinary work to validate against outcomes that actually matter. More costly are value models trained to predict a conversation's eventual outcome from a partial trace, then used to densify the reward. The long-horizon objective still has to be specified upstream; potential-based shaping \cite{ng1999policy} helps with credit assignment, not with specification, and is useful because it densifies a sparse signal without introducing a new optimum.

The most expensive is direct trajectory comparison, which includes preferring conversations that end in lower regret or greater clarity over those that end in \textit{``I wish I'd seen it sooner.''} This sidesteps fine-grained credit assignment but requires outcome labels to be gathered over time. These signals belong in multi-objective, constrained post-training, where helpfulness, support, truthfulness, uncertainty calibration, and agency are distinct targets, and failure modes such as unjustified agreement, escalation encouragement, and manipulation are explicitly bounded rather than left to emerge. Their relative weight should shift with context and the user's chosen role, rather than defaulting to a single supportive persona \cite{marks2026persona, lu2026assistant}.

\subsection{Product design affordances}
Because role mismatch is partly a product-design failure (Section~\ref{subsec:role_confusion}), well-being alignment also needs UI and interaction affordances that put user agency into practice \cite{schneider2018empowerment}. The interaction contract should be explicit.

Systems should offer clear mode selection (concierge/collaborator/coach) with plain-language descriptions of what changes, including when the model will challenge, how it handles uncertainty, and how it treats risky requests. When the model does push back, it should provide a brief, non-moralizing rationale tied to the user-stated goals, evidence, and uncertainty, as well as the relevant tension. Product affordances should also support graduated interventions (nudge $\rightarrow$ challenge $\rightarrow$ escalation suggestions in higher-stakes cases) rather than a brittle comply-versus-refuse policy \cite{hofman2023sports}. 

In addition, well-being modes should monitor trajectory drift, not only isolated policy violations. The system could flag patterns such as repeated reassurance-seeking, escalating certainty around weak evidence, increasing distress, sleep or work disruption, or dependence on the assistant for ordinary decisions \cite{kirk2025neural, xie2023friend}. The response need not be a hard refusal: the assistant can slow the interaction, name the pattern, ask whether the user wants a different mode, and introduce grounding prompts before escalating to stronger suggestions or external support. A coach mode should therefore differ in behavior across a conversation, not just in tone.

\subsection{Governance}
Providers that market a well-being mode should state what the mode is optimized for and how they evaluate it \cite{cath2018governing, taeihagh2021governance}. At minimum, providers should disclose what each mode is optimized for (e.g., engagement, immediate satisfaction, or longer-horizon outcomes) so users and auditors can read its behavior against the intended objective rather than marketing claims. In parallel, developers should publish model-card-style summaries that make these behaviors auditable over time \cite{mitchell2019model}, including endorsement and disagreement calibration, and update drift (whether a new release becomes systematically more affirming or more withholding in the same scenarios).

\section{Alternative Views}

\subsection{AV1: General-purpose preference optimization is the right default}
A credible view is that RLHF-style preference optimization remains the best default because it produces systems that are broadly usable, non-confrontational, and aligned with what most users immediately want. From this perspective, making ``long-horizon well-being'' a primary objective risks creating assistants that feel intrusive or moralizing, reducing adoption and increasing user backlash.

Our position does not require replacing the default. Rather, we argue that, as LLMs are used for socioemotional guidance, there should be at least one widely accessible, opt-in mode explicitly optimized and evaluated for longer-horizon outcomes, with transparent role contracts and auditability. This preserves usability for productivity-first use while addressing predictable failure modes in well-being contexts.

\subsection{AV2: Long-horizon outcome measurement is infeasible or privacy-risky}
A second view is that longitudinal well-being measurement is too noisy and too privacy-sensitive to serve as a training target. Follow-ups can be biased, sparse, and domain-specific, and attempts to operationalize them at scale may incentivize intrusive data collection or surveillance.

We agree that measurement is difficult and must be privacy-preserving by design. The current alternative is optimization against dense proxies like engagement and retention, which can themselves shape long-run outcomes while remaining unaudited. Our call is for opt-in, minimal, privacy-preserving outcome signals and explicit evaluation protocols that make these tradeoffs explicit and governable.

\subsection{AV3: Well-being alignment risks paternalism and value imposition}
A third alternative view is that optimizing LLMs for well-being inevitably embeds contested values and invites paternalism. Even if framed as coaching, pushback can become coercive, and product design can steer users toward a normative interaction style.

This concern is real, and our proposal answers it with explicit user choice, reversibility, and transparency. Users should select relational modes, see brief rationales for pushback tied to stated goals and evidence standards, and retain the ability to override or downgrade the mode. The goal is to make objectives explicit and auditable, and to keep empathic support distinct from epistemic or normative endorsement, not to enforce a single moral doctrine.

Opt-in design does not remove all paternalism, because defaults, labels, and interface framing can still steer users. These choices should themselves be audited.

\subsection{AV4: Product-layer design is sufficient; changing training is unnecessary}
A further alternative view is that well-being failures should be addressed primarily at the product layer, while keeping the base model general-purpose. This approach is appealing because it is faster to iterate and easier to govern than changing training objectives.

Product affordances help, but they are not enough on their own when the underlying model is optimized for short-horizon approval. If training remains preference-optimized, product scaffolding will continually conflict with the model's incentives and may be eroded over time by deployment optimization. Stable well-being behavior requires alignment across evaluation, objectives, and product affordances rather than relying solely on UI.

\section{Conclusion}
As LLMs take on socioemotional and life-guidance roles, next-turn approval becomes an increasingly poor proxy for success. We showed how preference optimization and deployment incentives reward behavior that looks helpful within a conversation while producing worse outcomes over time. We then laid out a design space around three tensions and used it to motivate specific recommendations spanning measurement, benchmarks, training objectives, product affordances, and governance (Table~\ref{tab:tensions_wellbeing_glance}).

This view complements work treating human-AI interaction as relational and socioaffective, and argue that alignment must account for socioaffective dynamics (e.g., \cite{kirk2025human, laukkonen2026positive}). For systems that increasingly advise people on how to live, the relevant test is what happens after the conversation ends.

\section*{Acknowledgements}
We thank the reviewers for their feedback. This publication was made possible through the support of Grant 63578 from the John Templeton Foundation. The opinions expressed in this publication are those of the author(s) and do not necessarily reflect the views of the John Templeton Foundation.

\bibliography{references}
\bibliographystyle{icml2026}


\end{document}